

\input phyzzx
\overfullrule=0pt
\tolerance=5000
\twelvepoint
%
\def\npb#1#2#3{{Nucl.\ Phys.} {\bf B#1} (19#2) #3.}
\def\plb#1#2#3{{Phys.\ Lett.} {\bf #1B} (19#2) #3.}

\def\prl#1#2#3{{Phys.\ Rev.\ Lett.} {\bf #1} (19#2) #3.}

\def\cmp#1#2#3{{Commun.\ Math. Phys.} {\bf #1} (19#2) #3.}
\def\p{\partial}                        
\def\frac#1#2{{#1 \over #2}}            
\def\vac{\ket 0}
\def\CG{{\cal G}}
\def\CC{{\cal C}}
\def\CP{{\cal P}}
\def\CJ{{\cal J}}

\def\bz{{\bar z}}
\def\bsi{{\bar \sigma}}
\def\bep{{\bar \epsilon}}
\def\bps{{\bar \psi}}
\def\brh{{\bar \rho}}
\def\ds{\raise.15ex\hbox{/}\kern-.57em\partial}
\def\Ds{\,\raise.15ex\hbox{/}\mkern-13.5mu D}
%
\pubnum{EFI-92-25}
\date{}
\titlepage
\title{Low-Energy Dynamics of Supersymmetric Solitons}
\vglue-.25in
\author{Jerome P. Gauntlett  \foot{e-mail: jerome@yukawa.uchicago.edu }}
\medskip
\address{\centerline{Enrico Fermi Institute}
\centerline{University of Chicago}
\centerline{5640 South Ellis Avenue}
\centerline{Chicago, IL 60637, USA}}
\bigskip
\abstract{
In bosonic field theories the low-energy scattering of solitons
that saturate Bogomol'nyi-type bounds can be approximated as
geodesic motion on the moduli space of static solutions.
In this paper we consider the analogous issue within the
context of supersymmetric field theories. We focus our study
on a class of
$N=2$ non-linear sigma models in $d=2+1$ based on an arbitrary
K\"ahler target manifold and their associated
soliton or ``lump" solutions.
Using a collective co-ordinate expansion, we construct
an effective action which, upon quantisation,
describes the low-energy dynamics of the lumps.
The effective action
is
an $N=2$ supersymmetric quantum mechanics action with the
target manifold being the moduli
space of static charge $N$ lump solutions of the sigma model.
The Hilbert space
of states of the effective theory consists of anti-holomorphic forms on
the moduli space. The normalisable elements
of the dolbeault cohomology classes $H^{(0,p)}$ of the moduli
space correspond to zero energy bound states and we argue that such states
correpond to bound states in the full quantum field theory of the sigma model
.}
\endpage
%
%
\REF\manton{N. Manton, \plb{110}{82}{54}}
\REF\atiyah{M. Atiyah and N. Hitchin, The Geometry and
Dynamics of Magnetic Monopoles, Princeton University Press, 1988.}
\REF\rubackt{P. Ruback, \npb{296}{88}{669}}
\REF\ward{R.S. Ward, \plb{158}{85}{424}}
\REF\gibbons{G.W. Gibbons and P. Ruback, \prl{57}{86}{1492}}
\REF\nick{G.W. Gibbons and N.S. Manton, \npb{274}{86}{183}}
\REF\bernd{B. Schroers, preprint DAMTP-91-05}
\REF\trevor{T.M. Samols, preprint DAMTP-91-13}
\REF\harvey{J. Harvey and A. Strominger, preprint EFI-91-30.}
\REF\wittent{E. Witten, \npb{202}{82}{253}}
\REF\montonnen{C. Montonnen and D. Olive, \plb{72}{77}{213}}
\REF\shellard{E.P.S. Shellard, Cosmic Strings:
the Current Status, 1988.}
\REF\jerome{J.P. Gauntlett and J. Harvey, in preparation.}
\REF\zak{A.M. Din and W.J. Zakzrewski, \npb{253}{85}{77}}
\REF\zakr{I. Stokoe and W.J. Zakzrewski, Z. Phys.
{\bf C34} (1987) 491.}
\REF\leese{R. Leese, \npb{344}{90}{33}}
\REF\ruback{P. Ruback, \cmp{116}{88}{645}}
\REF\zumino{B. Zumino, \plb{87}{79}{203}}
\REF\alvarez{L. Alvarez-Gaum\'e and D.Z. Freedman,
\cmp{80}{81}{443}}
\REF\witten{E. Witten and D. Olive, \plb{78}{78}{97}}
\REF\jeromet{J.A. de Azcarraga, J.P. Gauntlett, J.M. Izquierdo and
P.K. Townsend, \prl{63}{89}{2443}}
\REF\zuminot{B. Zumino, \plb{69}{77}{369}}
\REF\claudia{J.P. Gauntlett and C. Yastremiz, Class. and Quantum Grav.
{\bf 7} (1990) 2089.}
\REF\alva{
L. Alvarez-Gaume, \cmp{90}{83}{161}}
\REF\friedan{D. Friedan and P. Windey,
\npb{235}{84}{395}}
\REF\raj{R. Rajaraman, Solitons and Instantons, North-Holland, 1987}

%
\chapter{Introduction}
The study of the low-energy scattering of
soliton-like objects in bosonic field theories was initiated by
the work of Manton on monopoles [\manton].
It was argued that the scattering of slowly moving BPS monopoles
could
be approximated as geodesic motion on the moduli space of static
solutions.
Atiyah and Hitchin then continued the programme by explicitly
constructing the
metric on the moduli space of two monopoles
and then employing it to construct geodesics (see [\atiyah]
and references therein).
One of the most striking conclusions is that in a head on
collision two monopoles
scatter at $90^0$.

These techniques have since been applied to a variety of other models,
including Abelian-Higgs
vortices in two dimensions [\rubackt], ``lumps"
of non-linear sigma models [\ward]
and extremal black holes [\gibbons].
A common feature of all these models is that
the static multi-soliton solutions saturate a Bogomol'nyi-type bound.
This implies that there is no net force between the
solitons,
an essential ingredient in the geodesic approximation.
The presence of the Bogomol'nyi bounds is a result of the fact that
the models have supersymmetric extensions. Topological charges
appear in the supersymmetry algebra from which one can
deduce the Bogomol'nyi bound.

Given this deep connection with supersymmetry,
it is natural to consider the scattering of solitons in the
supersymmetric theory itself.
Since the fermionic equations of motion of the supersymmetric
theory are
trivially satisfied by setting all the fermions to zero,
the static soliton solutions of the bosonic theory continue to
be solutions in the supersymmetric theory.
However,
the inclusion of the fermions dramatically changes the quantum theory.
Now there will be fermionic zero modes in addition to the bosonic
zero modes in the fluctuations
about the classical soliton solution.
After quantisation, the fermionic zero modes imply that there
is a multiplet of soliton states degenerate in energy
(assuming that the supersymmetry is not dynamically broken).

This illustrates a crucial difference in principle between the
scattering of solitons in the bosonic and supersymmetric theories.
In the bosonic theory the geodesic approximation is developed
by viewing the time evolution of the classical solitons as that
of a fictitious particle moving in an infinite dimensional
configuration space, the space of time independent, finite energy
field configurations. That is,
the analysis is purely classical (although the quantum case can,
of course,
also be considered [\nick,\bernd,\trevor]).
In the supersymmetric case the r\^ole of the
fermionic zero modes is only manifest after quantisation.
To discuss the scattering of solitons in this case,
it is thus necessary
to construct an effective action which, upon quantisation,
describes the low energy dynamics of the system. In this paper
we construct such an action for a particular class of models
by employing a collective co-ordinate expansion.

In the bosonic case, the collective co-ordinate expansion is
essentially an
equivalent way to develop the geodesic approximation.
The fluctuations about the soliton solution contain zero modes
that determine the low-energy dynamics.
For each zero mode one must introduce a collective co-ordinate. The
collective co-ordinates are the arbritrary parameters or moduli that
a general static soliton solution depends on i.e. they can be considered
as providing
co-ordinates on the moduli space of static solutions. More physically,
they can be interpreted as corresponding
to the positions and ``charges" of
the solitons.
By allowing the collective co-ordinates to depend on time
and ignoring the
contributions of the non-zero modes one constructs an ansatz
for the low-energy fields. After
substituting this ansatz into the action one
obtains an effective action
describing the low-energy dynamics of the system.
The effective action is that of a free particle propagating
on the moduli space of static solutions, with the metric being
naturally induced from the kinetic energy functional of the
parent field theory. The solutions to the
equations of motion of the effective action
are simply the geodesics on the moduli space.

If the
exact soliton solutions are explicitly known,
this procedure can be carried out directly.
This is the approach employed in the
study of the scattering of the lumps of bosonic sigma models
[\ward].
For
most models, however, less is known about the soliton solutions.
Nevertheless, in some special cases, such as
the scattering of two solitons (the case of most interest),
the geometry of the moduli
space can still be derived using indirect methods
relying on identifying the various symmetries of the model.
This approach has successfully been used in the monopole system
[\atiyah].

In the supersymmetric case
we can use a blend of techniques.
It is a generic phenomenon that the
solutions saturating the Bogomol'nyi
bound, i.e. the static soliton solutions,
only break half of the supersymmetry. We will show how the unbroken
supersymmetry pairs the bosonic zero modes with the fermionic
zero modes. Using this
we can  formally perform the collective co-ordinate
expansion for the fermions and bosons. After identifying the
geometric structures of the moduli space we can then utilise any of
the results of the bosonic case that have been obtained either
directly or indirectly. We will see
that this naturally leads to an effective
action which is a kind of supersymmetric quantum mechanics on the moduli
space.
We note that a similar
method was employed in [\harvey] to show that the low-energy
scattering of axions with 5-branes in superstring theory is related to the
Donaldson polynomial. This connection between geometry and low-energy
supersymmetric dynamics
was one of the inspirations for this work.

We note that the procedure outlined above for deriving the
low energy effective action involves neglecting all of the
modes apart from the zero modes. Upon quantisation this is equivalent to
assuming that the low energy dynamics may be described by a wave
function of field configurations in
which all but the zero modes are in the
ground state. One may wonder about the validity of this approximation.
In particular this approximation seems most vulnerable for the situations
where the field theory has
a continuous spectrum
(e.g. monopoles, sigma model lumps).
In these cases one may ask whether or not solitons with fermionic zero
modes excited are stable against the emmission of soft
``photinos". If supersymmetry is not
dynamically broken then the spectrum
of soliton states will be exactly degenerate which implies the
stability of the states.
In some cases, including the models considered in this paper, the
issue of supersymmetry breaking can be decided by calculating the
Witten index $\rm Tr(-1)^F$ or one of its variants [\wittent].

Our approach
is to assume that at low enough energy all radiation
processes are significantly
supressed and that the effective action just depending on the zero modes
is a good approximation. The range of validity of this approximation
will depend on the detailed dynamics of the model considered.
Of course for models such as the Abelian Higgs
model for which there is a mass gap
we expect the effective action to be
a particularly
robust approximation to the true dynamics.

An important class of states of the effective action are the normalisable
wave functions with lowest energy. These correspond to bound states of
the full quantum field theory as there are no other states with lower
energy into which they could decay.
We will see that the normalisable states of the effective action
with zero energy
are given by normalisable harmonic forms on the moduli space.

There is a variety of interesting supersymmetric models to consider.
It would be extremely interesting to investigate the Montonnen and
Olive
conjecture [\montonnen] concerning the self-duality
of $N=4$ super Yang-Mills under the interchange of magnetic and electric
charges, along with
the monople and elementary particle sectors of the spectrum
and the interchange of strong and weak coupling.
Presumably some insight into this conjecture could be
attained by studying the
low-energy scattering of monopoles. Another interesting case
to consider would be supersymmetric vortices.
It is known that in a head on collision bosonic vortices scatter
at $90^0$ at least in the abelian-Higgs model [\rubackt]. It has been
further shown that this provides a plausibility argument
as to why cosmic strings intercommute [\shellard], a crucial
property in cosmic string based cosmology.
It is natural to wonder how the presence of fermionic zero modes
could alter these phenomena.

In this paper we study the low energy dynamics of the lumps
of $N=2$ non-linear sigma models in d=$2+1$ based on an arbitrary
K\"ahler target manifold. Such sigma models
have been studied for a variety of reasons. In $d=2+1$ the non-linear
sigma model is not a renormalisable quantum field theory so it should
be viewed as describing some low-energy effective behaviour in planar
physics with an in built cutoff. Our main motivation
for studying these models is that they do not have any of the additional
complications that the gauge invariance introduces in the models
mentioned above.
We will return to a study of those models in the
future [\jerome].

The Witten index of the $CP^N$ sigma models is given by $Tr(-1)^F=N+1$
[\wittent].
Thus supersymmetry is not dynamically broken for these models and this
ensures the stability of the lump states with fermionic zero modes
excited, despite the fact that the system has a continuous spectrum.
The non-renormalisability of the model is not relevant for our discussions
as we are only interested in the very low-energy dynamics.
There is one feature of sigma model solitons that is not shared by other
models:
the number of normalisable
zero modes is smaller than the
dimension of the moduli space. We will discuss
this point further in the text and see that it brings
in some complications in the
analysis of the quantum theory.
Despite this latter point, we believe
that this model is a good laboratory to both investigate
most conceptual issues and to develop some techniques for
discussing the low-energy dynamics
of supersymmetric solitons in general.

The classical
scattering of lumps in the bosonic $CP^1$ and $CP^N$ sigma-models
has been studied by various authors
[\ward,\zak-\leese]. The geometry of the moduli space for
an arbitrary K\"ahler target was
elucidated by Ruback [\ruback] and his results will be important in the
following. After first reviewing the
bosonic case and introducing some
notation in section 2, we will then
discuss the supersymmetric sigma models
in section 3. We discuss how the soliton solutions break half of the
supersymmetry and
show how the unbroken supersymmetry
pairs the bosonic and fermionic zero modes. The collective co-ordinate
expansion will then lead to an effective supersymmetric quantum mechanics
action describing the
low-energy dynamics of the solitons.
The quantisation of the effective action is discussed in section 4.
We show that the Hilbert space of states is given by anti-holomorphic
forms on the moduli space of static solutions. We also show that
the zero energy
bound states are given by normalisable elements of the
cohomology classes $H^{(0,p)}$
of the moduli space and we argue that they correspond to bound
lump states in the full quantum field theory of the sigma model.
In an appendix we
discuss how the operator ordering in the quantum mechanics of the
effective action
is related to the operator ordering in the vaccuum sector
of the parent sigma model. Section 5 contains some conclusions.

\chapter{Bosonic Lumps}
Consider a bosonic non-linear sigma model with an
arbitrary  K\"ahler
target manifold M, described by the action
$$
S=-{1\over 2}\int d^3xg_{ij}(\phi(x))\partial_m\phi^i\partial
_n\phi^j\eta^{mn}\eqn\sig
$$
where $\phi^i$ and $g_{ij},\ \ i,j=1...2n$ are
co-ordinates and the metric on M, respectively.
To obtain the static soliton or lump solutions to the equations of motion,
we first rewrite the action in the form
$$
S=\int dt(K-V)\qquad,\eqn\sigt
$$
where the kinetic energy functional is given
by
$$
K={1\over 2}\int d^2x g_{ij}\dot\phi^i\dot\phi^j
\eqn\kin
$$
and the potential energy functional is
$$
V={1\over 2}\int d^2x g_{ij}\partial_\mu\phi^i\partial
_\mu\phi^j
\qquad,\eqn\pot
$$
where $\mu=1,2$. The total conserved energy is $E=K+V$.

The configuration space of the system, $\CC$, is the set of all finite energy
maps from $R^2$ to M. The action can then be viewed as describing
the motion of a fictitious particle moving on the infinite dimensional
space $\CC$ under the influence of the potential \pot . In this spirit,
the kinetic energy functional naturally defines a metric on $\CC$.
Specifically, if we let $\dot\chi^i(x),\,\dot\rho^i(x)$
be two tangent
vectors above the point $\phi^i(x)\in\CC$, then the metric $\tilde g$
is given by
$$
\tilde g(\dot\chi,\dot\rho)\equiv
\int d^2x g_{ij}(\phi(x))\dot\chi^i(x)
\dot\rho^j(x)\qquad.
\eqn\metc
$$
Similarly, the complex structure $J$ on the K\"ahler target
induces a natural complex structure $\tilde J$ on \CC\ via
$$
\left[\tilde J\dot\chi\right]^i(x)\equiv
{J^i}_j(\phi(x))\dot\chi^j(x)
\qquad.\eqn\compc
$$
The geometry of \CC\ was described in detail in [\ruback]
and,
assuming that \CC\ is at least an incomplete manifold (as we also will),
it was shown that $(\CC,\tilde g,\tilde J)$ is in fact
K\"ahler.

Static solutions to the equations of
motion of \sig\ are obtained by minimising the static energy
functional $V$.
As is well known, \pot\ can be recast in the form
$$
V={1\over 4}\int d^2x D^{\pm}_\mu\phi^i D^{\pm}_\mu\phi^j g_{ij}
\mp T
\qquad,\eqn\pott
$$
where
$$
D^{\pm}_\mu\phi^i=\partial_\mu\phi^i\pm {J^i}_j\epsilon_{\mu\nu}
\partial_\nu\phi^j
\eqn\hol
$$
and
$$
T=\int_{R^2}\phi^*(\omega)
=\int d^2 x g_{ik}{J^k}_j\partial_\mu\phi^i\partial_\nu
\phi^j\epsilon_{\mu\nu}
\eqn\kal
$$
is the  integral of the pull back of the K\"ahler form on M. Since
the K\"ahler form defines a non-trivial element of the second
cohomology class $H^2(M,R)$, $T$ is a
topological invariant. (For $M=CP^N$, it is,
up to normalisation, an
integer labelling the 2-d instanton number;
the instantons in d=2 become the static lumps in d=2+1).
The configuration space \CC\ is partitioned into different topological
sectors labelled by the topological charge $T$.

{}From \pott\ we can deduce the Bogomol'nyi bound on the static energy
functional
$$
E\ge \vert T\vert\qquad.\eqn\bog
$$
Within each topological class of maps labelled by the
topological charge $T$, the energy is minimised
when the Bogomol'nyi bound \bog\ is saturated.
Thus the static soliton solutions to the equations of motion
are obtained by solving the first order
differential equations
$D^\pm_\mu\phi^i=0$.
By switching to complex co-ordinates
it is easily seen that this corresponds to (anti-)holomorphic maps:
$\p_{\bar z}\phi^\alpha=0\ (\p_z\phi^\alpha=0)$. In the
following we will restrict our considerations to the case
of holomorphic maps; the analysis can be trivially extended
to the case of anti-holomorphic maps.

In general, the holomorphic maps will
depend on a continuous set of parameters, or moduli.
These moduli can
be interpreted as the positions and ``charges" of the static lumps.
The moduli space $M_N\subset \CC$ is defined as the set of all
holomorphic maps within a given topological class $T=N$. Clearly
the dimension of the moduli space is determined by the number of moduli.
The existence of exact static charge $N$-lump configurations
relies on the Bogomol'nyi bound being saturated.
This can be interpreted as saying
that there is no net force between the (static) lumps.
This is entirely analogous to the case of BPS monopoles.
Thus, following the work by Manton [\manton],
Ward  suggested in [\ward] that
the scattering of slowly moving lumps could be approximated
by assuming that the evolution is adiabatic in the space of
static soultions. That is, the time evolution is determined
by geodesic motion in the moduli space $M_N$.

The low energy scattering problem is thus reduced to determining
the geometry of $M_N$ and calculating its geodesics.
We first introduce some notation. We denote
the most general N-lump solution by $\phi^i_o(x,X)$
where the moduli $X^a, \ a=1,...,2k=dim(M_N)$, can be considered
as co-ordinates on $M_N$. The maps $\phi^i_o(x,X)$
can be considered as providing alternative co-ordinates
on $M_N$.
By letting the moduli $X$ depend on a parameter $t$, which we
interpret as time, we obtain the tangent vectors on $M_N$ by
differentiation:
$$
\dot\phi^i_o(x,X(t))\vert_{t=0}=\left.{\p\phi^i_o\over \p X^a}
\right\vert_{t=0}\dot X^a\equiv
\delta_a\phi^i\dot X^a\qquad.\eqn\tv
$$
Geometrically, $\delta_a\phi^i$ is the matrix corresponding to
a change of co-ordinates. Physically, for each $a$
they are simply the zero
modes in the fluctuations about the solution $\phi^i_o(x)\equiv
\phi^i_o(x,X(0))$.
This
can be seen by noting that, by definition,
$V(\phi_o^i(x,X(t))=V(\phi_o^i(x))$.

A metric $\CG$ on $M_N$ is naturally induced by the restriction of the
metric \metc\ on \CC :
$$
\CG(
 \dot\phi_o,\dot\phi_o)\equiv
 \CG_{ab}(\phi_o)\dot X^a\dot X^b\qquad,
\eqn\met
$$
where
$$
\CG_{ab}=\int d^2x g_{ij}(\phi_o(x))\delta_a\phi^i\delta_b\phi^j
\eqn\metm
$$
is the representation in the co-ordinates $\{X^a\}$.
We note that the physical assumption, due to finiteness of
energy, of only considering
zero modes with finite norm (i.e. $\CG_{ab}$ finite)
is equivalent to only considering
tangent vectors with finite length.
Thus, although it might seem
that an arbitrary holomorphic perturbation of a holomorphic map
within a given topological class
would be a zero mode, only
a finite number of them will have finite norm.
A feature peculiar to sigma model solitons is that
the number of normalisable zero modes is less than the
dimension of $M_N$ [\ward]. Thus the time evolution of
slowly moving lumps will be geodesic motion on some submanifold
$\tilde M_N\subset M_N$, defined by fixing all the collective co-ordinates
corresponding to zero modes with infinite norm to be constant.

It was shown in [\ruback] that if
$\dot \phi_o^i$ is a tangent vector to $M_N$ then so is
$[\tilde J\dot\phi_o]$. Thus the restriction of the complex
structure \compc\ on
\CC\ provides a natural complex structure on the moduli space.
In the co-ordinates $\{X^a\}$ it has the following form:
$$
{\CJ^a}_b=\CG^{ac}\int d^2x{J^i}_jg_{ik}\delta_c\phi^k
\delta_b\phi^j\qquad.
\eqn\com
$$
The property $\CJ^2=-{\bf 1}$ is most easily verified using the co-ordinates
$\phi^i_o(x,X)$. By a generalisation of the argument in [\ruback], it
is clear that \com\
provides a complex structure on $\tilde M_N$ by
restricting the zero modes to have finite norm.
Thus $(\tilde M_N, \CG, \CJ)$ is a K\"ahler manifold.

We conclude this section with a brief discussion of the symmetry groups
of the various spaces we have been discussing. From \metc\ it is clear
that the isometry group of
$(\CC,\tilde g)$ is given by the product of the isometry group of $R^2$ with
the isometry group of the target manifold $M$. Considering \compc\ we
conclude that the subgroup given by the product of the isometry group of
$R^2$ with the group of holomorphic isometries on $M$ are holomorphic
isometries of $(\CC,\tilde g,\tilde J)$. The holomorphic isometries
of $(\CC,\tilde g,\tilde J)$ are holomorphic isometries of $(M_N,\CG,\CJ)$
since they both act covariantly on the static equations of motion and they
preserve the topological charge \kal. The combination of these isometries
that do not shift the collective co-ordinates held fixed in defining
$\tilde M_N$ provide the isometries of this space.

\chapter{Supersymmetric Lumps}
The sigma model \sig\ has a supersymmetric extension given
by
$$
S=-{1\over 2}\int d^3x\left\{g_{ij}\partial_m\phi^i\partial
_n\phi^j\eta^{mn}
+i\bps^i\Ds\psi^j g_{ij}+{1\over 6}R_{ijkl}\bps^i\psi^k
\bps^j\psi^l\right\}\eqn\susysig
$$
where $\psi^i$ is a set of two component Majorana SL(2,R) spinors that
transform as a vector on the target manifold M. The covariant
derivative in \susysig\ is defined using the pullback of
the connection on M:
$$
D_m\psi^i=\partial_m\psi^i+{\Gamma^i}_{jk}\partial_m\phi^j
\psi^k\qquad.\eqn\covd
$$
Our conventions for the gamma matrices are as follows:
${(\gamma^1)^\alpha}_\beta={(\sigma^1)^\alpha}_\beta,\
{(\gamma^2)^\alpha}_\beta={(\sigma^3)^\alpha}_\beta,\
{(\gamma^0)^\alpha}_\beta=C_{\alpha\beta}=
{(i\sigma^2)^\alpha}_\beta$, where $C$ is the charge conjugation matrix.
The Dirac adjoint is thus given by ${\bar\psi}^i={\psi^i}^\dagger\gamma^0=
{\psi^i}^TC$.

The action is invariant under the supersymmetry transformation
$$
\eqalign{
\delta\phi^i&=i\bep^1\psi^i\cr
\delta\psi^i&=\ds\phi^i\epsilon^1-i{\Gamma^i}_{jk}\bps^j\epsilon^1
\psi^k\cr}\eqn\onesusy
$$
where $\epsilon^1$ is a constant anticommuting Majorana spinor.
Additional supersymmetries require a reduction of the holonomy
of M [\zumino,\alvarez]. In the case at hand, M is K\"ahler and
hence there is an additional supersymmetry given by
$$
\eqalign{
\delta\phi^i&=i\bep^2{J^i}_j\psi^j\cr
\delta\psi^i&=-{J^i}_j\ds\phi^j\epsilon^2
-i{\Gamma^i}_{jk}{J^k}_l\bep^2\psi^l
\psi^j\cr}\eqn\twosusy
$$
where $\epsilon^2$ is a second constant anticommuting Majorana spinor.

The fermionic equations of motion of \susysig\ are trivially
satisfied by setting all of the fermions to zero. Thus the static
soliton solutions of the bosonic theory continue to be solitons in
the supersymmetric theory.
In the following we will call the solution $\p_\bz\phi_o^\alpha=\psi_o
^\alpha=0$
the supersymmetric solution, where $\alpha=1,...,n$ are holomorphic
co-ordinates on the target.

The supersymmetric solutions break
half of the supersymmetry. By this we mean that of the four
dimensional space spanned by the supersymmetry parameters, the
solution is left invariant by a two dimensional subspace. To see
this we use holomorphic co-ordinates
and introduce the following hermitean projection
operator in the spinor space:
$$\Gamma={1\over4}\gamma^z\gamma^\bz,
\eqn\proj
$$
satisfying
$$
\Gamma^2=\Gamma,\qquad \Gamma+\Gamma^*=1,\qquad \Gamma\Gamma^*=0
\eqn\projp
$$
where $\gamma^z\equiv\gamma^1+i\gamma^2$.

After redefining the supersymmetry paramaters via
$$
\chi\equiv\epsilon^1+i\epsilon^2
\qquad \Gamma\chi\equiv\rho
\qquad \Gamma^*\chi\equiv\sigma
\eqn\spdec
$$
the $N=2$ supersymmetry transformations \onesusy, \twosusy\
acting on a static bosonic holomorphic map $\phi^\alpha_o$
are given by
$$
\eqalign{
\delta_\sigma\phi^\alpha&=
i\bsi^*\psi^\alpha\cr
\delta_\sigma\psi^\alpha&=-i{\Gamma^\alpha}_
{\beta\gamma}|_{\phi_o}\bsi^*\psi^\beta\psi^\gamma\cr
}\eqn\ususy
$$
and
$$
\eqalign{
\delta_\rho\phi^\alpha&=0\cr
\delta_\rho\psi^\alpha&=\gamma^z\partial_z\phi_o^\alpha\rho^*
-i{\Gamma^\alpha}_{\beta\gamma}|_{\phi_o}\brh^*\psi^\beta\psi^\gamma\cr
}\eqn\bsusy
$$
where we have used the fact that for a K\"ahler manifold
the Christoffel symbol
is pure in its holomorphic indices.
When the fermions are set to zero, $\delta_\sigma$ is
the unbroken supersymmetry leaving the supersymmetric solution invariant and
$\delta_\rho$ is the broken supersymmetry.

This partial breaking of supersymmetry is a generic feature
of supersymmetric field theories admitting topologically
non-trivial solutions.
It was first noticed by Witten and Olive [\witten] and is best
understood by showing that the algebra of supersymmetry charges
are modified by topological charges (see [\jeromet] for a model
independent discussion of this). For the N=2 supersymmetric
sigma model it was shown in [\ruback] that the supersymmetry
algebra is given by
$$
\left\{Q^I_\alpha,Q^J_\beta\right\}=
\delta^{IJ}P^m(\gamma_m)_{\alpha\beta}
+T\epsilon^{IJ}C_{\alpha\beta}
\eqn\susyalg
$$
where $T$ is the topological charge \kal. Furthermore, from
\susyalg\ we can deduce the Bogomol'nyi bound \bog\ and that the
bound is saturated iff the solution breaks half of the
supersymmetry.

We now turn to a discussion of the zero modes in the fluctuations
about the supersymmetric solution. The bosonic zero modes are
exactly the same as for the bosonic sigma model since after setting
the fermions to zero in the equations of motion of the supersymmetric
model one obtains the same equations of motion as in the bosonic model.
The fermionic zero modes are normalisable c-number solutions
to the Dirac equation in the presence of the soliton background.
A metric on the space of fermion zero modes is induced by the fermion
kinetic term in the action \susysig:
$$\CG'_{a'b'}=
\int d^2x g_{ij}(\phi_o(x))(\psi^i_{a'})^T\psi^j_{b'}
\eqn\fermet
$$
where we have denoted the fermionic zero modes by
$\psi^i_{a'}$.
Restricting our considerations to normalisable zero modes
is equivalent to demanding that $\CG'_{a'b'}$ has finite entries.

Two normalisable fermionic zero modes are immediately obtained from
the broken supersymmetry.
The fermionic part of the supersymmetry
transformation of the supersymmetric solution is given by
$$
\psi^\alpha=\gamma^z\p_z\phi_o^\alpha\rho^*
\qquad,\eqn\ozm
$$
and it is straightforward to verify that it
satisfies $\Ds(\psi^\alpha)=0$. Clearly these
zero modes satisfy $\Gamma\psi^\alpha=\psi^\alpha$.
These modes can be interpreted as the Goldstone
modes of the broken supersymmetry.

Following an argument by Zumino [\zuminot] in the
context of instantons, we will now show that the unbroken
supersymmetry pairs {\it all} of the bosonic and fermionic
zero modes. This pairing will be crucial to the derivation
of the effective action.
After imposing the following restrictions
$$
\partial_\bz\phi^\alpha=0
\qquad\psi^\alpha=\Gamma\psi^\alpha
\qquad,\eqn\eomans
$$
the equations of motion for time independent
configurations take the form
$$
\partial_\bz\phi^\alpha=
\partial_\bz\psi^\alpha=0
\qquad.\eqn\eomp
$$
Returning to the supersymmetric solution
$\partial_\bz\phi_o^\alpha=\psi^\alpha_o=0$,
this seems to indicate that any holomorphic bosonic
or fermionic perturbations are zero modes. In particular, it is
surprising that the Dirac equation does not seem to depend on the background.
We discussed in detail
in the last section that the normalisable
bosonic zero modes are restricted
by requiring that the metric on the moduli space
\metm\ has finite entries. The fermionic zero modes
are similarly restricted
by demanding
that the c-number
holomorphic perturbations be normalisable
(finite entries in \fermet). We note that this also resolves the apparent
paradox of the background independence of the Dirac equation in \eomp.

The linearly realised
supersymmetry now
reads
$$
\delta_\sigma\phi^\alpha=i\bsi^*\psi^\alpha
\qquad\delta_\sigma\psi^\alpha=0
\qquad.\eqn\linsusy
$$
If we let $\psi^\alpha$ be a normalisable fermionic zero mode and we
let $\sigma$ also be a c-number spinor, it is clear that
\linsusy\
generates a normalisable
bosonic zero mode.
This would seem to imply that for each
fermionic zero mode there are two bosonic zero modes. However,
because the supersymmetry
algebra \susyalg\ has an $SO(2)$ automorphism group,
these bosonic zero modes are not independent.
Using this we can invert \linsusy\ to supersymmetrically pair
the zero modes via
$$
\psi_p^\alpha=\delta_p\phi^\alpha\epsilon\eqn\pairing
$$
where $\epsilon$ is a c-number spinor satisfying
$\Gamma\epsilon
=\epsilon$, $\epsilon^\dagger\epsilon=1$ and here and in the following
$p,q,r$ are holomorphic
co-ordinates on $\tilde M_N$.
Thus for each normalisable bosonic zero
mode there is in fact one normalisable fermionic zero mode.
We further note that \pairing\ implies that the
metrics
\metm\ and \fermet\ are equal, consistent with the unbroken supersymmetry.

At this stage we have shown that the fermionic zero modes satisfying
$\Gamma\psi^\alpha=\psi^\alpha$ are paired with the bosonic
zero modes. In principle there could other fermionic zero modes
and one would need a kind of index argument to determine whether or
not they are present.
Although we do not have a general proof that there are not
additional fermionic zero modes, we do not expect them.
In all supersymmetric models that we know of where there is a corresponding
index theorem, a simple counting argument shows that all of the
bosonic and fermionic zero modes are paired by the unbroken supersymmetry.

The construction of the
effective action describing the low energy dynamics now
proceeds by a collective co-ordinate expansion.
For each normalisable zero mode, we introduce a collective co-ordinate.
For the bosonic zero modes this amounts to allowing the moduli
associated
with finite norm zero modes to depend on time
For the fermionic zero modes we use \pairing\
to introduce the collective co-ordinates in a way that preserves
the unbroken supersymmetry. Specifically, we are led
to the following low-energy ansatz for the time varying fields
$$
\eqalign{
\phi^\alpha(t,z)&=\phi_o^\alpha(z,X^p(t))+\dots\cr
\psi^\alpha(t,z)&=\delta_p\phi^\alpha\epsilon^p(t)+\dots\cr
}
\eqn\ansatz
$$
where $\epsilon^p$ is now a Grassmann odd spinor satisfying
$\Gamma\epsilon^p=\epsilon^p$ and the neglected terms correspond
to non-zero modes.
This expansion corresponds to a change of variables
from $\phi$ and $\psi$ to their infinite mode expansions and
the low-energy approximation is imposed by neglecting
all but the zero modes.
After substituting the ansatz \ansatz\ into the action \susysig\
and exploiting
the fact that the moduli space $\tilde M_N$
is K\"ahler, we obtain the following effective
action describing the low-energy dynamics:
$$
S_{\rm eff}=\int dt\CG_{p\bar q}\left\{{\dot X}^p
\dot{\bar X}^{\bar q}
+i\lambda^p D_t\lambda^{\bar q}\right\}
\eqn\seff
$$
where $\epsilon^p\equiv{1\over \sqrt 2}\left(
\matrix{\lambda^p\cr -i\lambda^p\cr}
\right)$
and
the covariant derivative is defined
using the pullback of the Christoffel connection $\gamma$
on the moduli space:
$$
D_t\lambda^{\bar p}=\dot\lambda^{\bar p}+ {\gamma^{\bar p}}_{
{\bar q}{\bar r}}
\dot{\bar X}^{\bar q}\lambda^{\bar r}
\qquad.\eqn\covdt
$$
The action can be recast into a slightly more familiar form
using real co-ordinates:
$$
S_{\rm eff}={1\over 2}\int dt\CG_{ab}\left\{{\dot X}^a
\dot{\bar X}^b
+i\lambda^a D_t\lambda^b\right\}\qquad.
\eqn\seffr
$$
This action is a variation of the usual supersymmetric
quantum mechanics action.
Usually, the fermions are two component
Majorana spinors that transform as real vectors on the target.
In the present case the $\lambda^a$ are a set of one component
Grassmann odd objects that transform as real vectors on the target.
This fact and the ansatz
\ansatz\ resolve the apparent contradiction discussed in the
conclusion of [\ruback].

The unbroken supersymmetry of the underlying
sigma model provides an N=2 supersymmetry of \seff\
i.e. there exist two one component anti-commuting
supercharges
\foot{On a general manifold the action \seffr\
is sometimes called an $N={1\over 2}$ supersymmetric quantum mechanics.
In the present context where we have two $N={1\over 2}$ supersymmetries
the nomenclature is clearly clumsy.}.
To show this we
define the unbroken supersymmetry parameter by
$\sigma={1\over \sqrt 2}
\left(\matrix{i\kappa\cr -\kappa\cr}\right)$,
satisfying $\Gamma^*\sigma=\sigma$, and using
\onesusy,\twosusy,\spdec\ and \ansatz\
we deduce that
$$
\eqalign{\delta X^p&=i\kappa\lambda^p\cr
\delta\lambda^p&=-\dot X^p\kappa^*\qquad.\cr}
\eqn\susy
$$
The supersymmetry charges can be derived using Noether's theorem and we
find
$$
\eqalign{
Q&=\CG_{p{\bar q}}{\dot X}^{\bar q}\lambda^p\cr
Q^*&=\CG_{p{\bar q}}{\dot X}^{p}\lambda^{\bar q}\qquad.\cr}
\eqn\charges
$$
Thus the effective action describing the low energy dynamics
of a set of N supersymmetric solitons is
given by an N=2 supersymmetric quantum mechanics with the moduli
space of N static solutions as a target manifold.

It will be useful in the following to know how the angular
momentum operator of the sigma model
depends on the zero modes. This could be derived
by first calculating the angular momentum operator of the sigma model and
then substituting \ansatz. A more direct way is the following. We first
define the two supersymmetry charges of the sigma model via
$$
\delta\phi^i=(i\bep^1Q^1+i\bep^2Q^2)\phi^i\qquad.
\eqn\th
$$
Using \onesusy\ and \twosusy\ a short calculation shows that up to a factor
the supersymmetry charge $Q$ in \charges\ corresponds to the
following linear combination of the components of the sigma model
supersymmetry charges defined in \th:
$$
Q={1\over 2}\left[Q^1_1+Q^2_2+i(Q^1_2-Q^2_1)\right]
\qquad.
\eqn\fr
$$
In [\claudia] it was shown that the d=2+1 superPoincare algebra
implies
$$
[J,Q]=-{1\over 2}Q\qquad [J,Q^*]={1\over 2}
\eqn\frog
$$
where $J\equiv M_{12}$ is the angular momentum generator in d=2+1.
After carrying out a canonical analysis of \seff\ we can construct
the zero mode contribution to $J$ by demanding that it satisfies \frog\
and commutes with the effective Hamiltonian. This is done in the next section.

Before concluding this section,
we note that the consistency condition for
the existence of the N=2 supersymmetric
quantum mechanics is that the moduli space of static solutions be
K\"ahler, as indeed it is. It is tempting to suggest that
this logic can be reversed; that the K\"ahler nature of the moduli space
can be deduced from the presence of the unbroken supersymmetry.
This connection could be a general feature of field theories with Bogomol'nyi
bounds, as they all have unbroken supersymmetry and the moduli spaces
of static solutions are K\"ahler and hyperK\"ahler in the situations
when we expect N=2 and N=4 supersymmetric quantum mechanics, respectively.
It is possible that this connection is just a formal one and not a
substitute for hard analysis.

\chapter{Quantisation of the Effective Action}
To further pursue the analysis of the low energy dynamics of the
supersymmetric lumps, one needs to quantize
the effective action \seff.
We will see that additional information
from the parent field theory is still required to resolve
some ambiguities in the quantisation procedure.
As we mentioned in the introduction, the
quantum mechanics of the effective action describes the dynamics of the
full field
theory assuming that all but the zero modes are in the ground state.

To simplify the discussion we will discuss the quantisation
in the context of the $CP^1$ sigma model.
The moduli space $M_N$ for this target
consists of all rational functions
of degree $N$. The most general one lump solution is
given by
$$
\phi_o=\alpha+\beta(z+\gamma)^{-1}\eqn\chgone
$$
where the moduli $\alpha,\beta,\gamma$ are arbitrary complex
numbers. Only the modular parameter
$\gamma$ is associated to a zero mode of
finite norm, so only $\gamma$ is allowed to be time
dependent in \ansatz. Fixing $\alpha$ and $\beta$, the reduced
moduli space $\tilde M_N$ for one
lump is two dimensional with a flat metric
and $\gamma$ can be
interpreted as the location of the soliton in the plane.
For this case $M_N$ is simply the orbit space of the isometries
induced by translations in $R^2$.

The effective action \seff\ is thus an N=2 supersymmetric
quantum mechanics based on $R^2$. We can rewrite it in
hamiltonian form as
$$
S_{eff}=\int dt \left\{P_a\dot X^a+{im\over 2}\lambda^a\dot\lambda^b\delta
_{ab}-{P^2\over 2m}\right\}\eqn\effone
$$
where $a,b=1,2$ and the mass of the lumps $m$ has been defined via
$\CG_{ab}=m\delta_{ab}$.
In complex co-ordinates the supersymmetry charges \charges\ are given by
$$
Q=\lambda P\qquad Q^*=\lambda^* P^*\qquad.
\eqn\susyone
$$
The non-vanishing
canonical quantum commutation relations can be
read off directly from \effone, viz
$$
[X^a,P_b]=i\delta^a_b\qquad \{\lambda,\lambda^*\}=m^{-1}
\eqn\comone
$$
where here and in the following we set $\hbar=1$.
Defining a state $\vac$ satisfying $\lambda\vac=0$, the Hilbert space consists
of two types of states: $\vac f(X)$ and $\lambda^*\vac g(X)$. The
momentum operator is realised on these states in the usual way: $P_a=-i
{\partial\over\partial X^a}$.
The Hamiltonian is constructed using the
supersymmetry algebra:
$$
H=\{Q,Q^*\}={P^2\over 2m}
\eqn\hamilo
$$

The operator given by
$$
S=-{1\over 2}\lambda\lambda^*
\eqn\J
$$
clearly commutes with the Hamiltonian \hamilo\
and satisfies \frog. Thus we can interpret it as the semiclassical spin
operator in the one-lump sector.
Noting that
$$
[S,\lambda]=-{1\over 2}\lambda\qquad [S,\lambda^*]={1\over 2}\lambda^*
\qquad.\eqn\qj
$$
we deduce that the spin of the two types of states differ by a half
and that the Hilbert space consists of a d=2+1 supermultiplet.
An alternative way to arrive at this conclusion would be to use the
arguments developed in
[\claudia] to show that the N=2 superparticle in d=2+1 is a more
accurate description of the dynamics of a single supersymmetric lump
since it posesses all of the appropriate symmetries: broken
symmetries are non-linearly realised and unbroken symmetries are linearly
realised. In fact \effone\ is a gauged fixed version of this superparticle
action. The quantisation of superparticle actions naturally leads to
supermultiplets.
A more sophisticated analysis is required to determine the exact spin
content of the spectrum. It is natural to expect that there is one
boson and one fermion state and that an anyonic supermultiplet could
be arrived at if one included a Hopf term in the sigma-model action.

We now turn to a discussion of the quantisation in the multi-lump
sectors. Because of the translation invariance of the underlying
filed theory, the moduli space $\tilde M_N$ will factorise into
a flat piece
and a non-trivial piece $\tilde M_N^0$ [\ward].
The flat piece corresponds to the centre of mass motion, the orbit space
of the translations,
and its quantisation is the same as for the one lump
case leading to a d=2+1 supermultiplet. The quantum states of \seff\
are then obtained as a tensor product of these states and those
coming from the quantisation of the non-trivial piece.
Thus the non-trivial aspect of the dynamics concerns
quantising on the non-trivial part of the moduli space so we now
turn to a discussion of this.

We want to quantise the action \seff\ on an arbitrary K\"ahler manifold
$\tilde M_N^0$.
The quantisation of the
action \seff\ on an arbitrary compact manifold has been considered
in [\alva,\friedan]. It was shown that a natural way of quantising
the system leads to a Hilbert space of states consisting of
spinors on the target space (assuming the target admits a spin
structure). In the present case, where the target space
is K\"ahler we will show that an equally natural quantisation procedure leads
to a Hilbert space of states consisting of the anti-holomorphic forms
on the target. The difference between the two quantisations is due
to an operator ordering ambiguity in the transition from classical
to quantum mechanics\foot{If the moduli space were a Calabi-Yau manifold
then these two quantisations would be equivalent.}.
The operator ordering can be determined by
demanding that the operator ordering of the zero modes is
consistent with the operator ordering of the parent field theory in the
vaccuum sector.
In the appendix we show that an operator ordering of the sigma-model
consistent with the ordering adopted in [\wittent] to calculate the
Witten index of the theory, leads to an ordering producing the
Hilbert space of anti-holomorphic froms on the target.

We now show how this quantisation procedure works.
The quantisation of \seff\ is facilitated
by introducing tangent space indices.
Since a K\"ahler manifold has $U(N)$ holonomy, we can introduce a basis
of one forms $\{e^A_p,e^{\bar A}_{\bar p}\}$ satisfying
$$
\CG_{p\bar q}=e^A_pe^{\bar B}_{\bar q}\delta_{A\bar B}
\eqn\frame
$$
with $(e^A_p)^*=e^{\bar A}_{\bar p}$. Furthermore the nonvanishing
components of the
spin connection can be chosen to take the form
$$
\eqalign{
{{\omega_p}^A}_B&=e^A_q({E^q_B,}_p+{\gamma^q}_{pr}E^r_B)
\qquad and \qquad c.c.\cr
{{\omega_{\bar p}}^A}_B&=e^A_q{E^q_B,}_{\bar p}\qquad \qquad\qquad
and \qquad c.c.\cr}
\eqn\spincon
$$
where we have introduced $E^p_A$ satisying $e^A_pE^p_B=\delta^A_B$,
$e^A_pE^q_A=\delta^q_p$ and $``c.c"$ denotes the complex conjugates.
Using these definitions we can rewrite the action \seff\ in the
form
$$
S=\int dt \left\{\CG_{p\bar q}\dot X^p\dot X^{\bar q}+i\lambda^A D_t\lambda
^{\bar B}\delta_{A\bar B}\right\}
\eqn\sefftangent
$$
where the covariant derivative is now defined using the pullback of
the spin connection from the target to the worldline.

The canonical momenta are given by
$$
\eqalign{
P_p&=\CG_{p{\bar q}}\dot X^{\bar q} +i\lambda^A\lambda^{\bar B}
\omega_{pA{\bar B}}\qquad and \qquad c.c.\cr
\CP_{A}&={\delta L\over \delta \lambda^{A}}= 0\cr
\CP_{\bar B}&={\delta L\over \delta \lambda^{\bar B}}=
-i\lambda^A\delta_{A{\bar B}}\qquad.\cr}
\eqn\mom
$$
The definition of the momenta conjugate to the fermionic
variables neccesarily contain second class constraints.
These can be eliminated by replacing the following
non-zero graded Poisson brackets
$$
\{X^p,P_q\}_{\rm pb}=\delta^p_q\qquad \{\lambda^p,\CP_q\}_{\rm pb}
=-\delta^p_q\qquad and \qquad c.c.
\eqn\pois
$$
by Dirac brackets.
To canonically quantise one then replaces the (graded) Dirac
brackets by (graded) commutators via $\{\ ,\ \}_{DB}\to -i[\ ,\ ]$.
A short calculation shows that the non-zero commutators are given by
$$
\eqalign{
[X^p,P_q]&=i\delta^p_q\qquad and \qquad c.c.\cr
\{\lambda^A,\lambda^{\bar B}\}&=\delta^{A\bar B}\qquad.\cr}
\eqn\com
$$
The holomorphic and anti-holomorphic components of the fermions
are thus raising and lowering opreators.
Defining a state
$\vac$ satisfying $\lambda^A\vac=0$ the Hilbert space consists
of the following states:
$$
\eqalign{
\vert f>
&=\lambda^{\bar A_1}\dots\lambda^{{\bar A}_p}\vac
{1\over p!}f_{{\bar A}_1\dots{\bar A}_p}(X)\cr
&=\lambda^{{\bar p}_1}\dots\lambda^{{\bar p}_p}\vac
{1\over p!}f_{{\bar p}_1\dots{\bar p}_p}(X)\qquad.\cr}
\eqn\states
$$
Acting on these states the bosonic momenta are realised in the usual way:
$P_p=-i{\p\over\p X^p}$ and
$P_{\bar p}=-i{\p\over\p X^{\bar p}}$.
Thus the Hilbert space can be identified with the space of
anti-holomorphic forms on the moduli space.

The natural inner product is given by the hermitean inner
product of the differential forms. If $\vert f>$ and $\vert g>$ are two
states corresponding to p-forms the inner product is given by
$$
<f\vert g>={1\over p!}\int dX d{\bar X}\sqrt{\CG}{\bar f}
^{{\bar p}_1\dots{\bar p}_p}
g_{{\bar p}_1\dots{\bar p}_p}\qquad.
\eqn\inner
$$
The inner product of two states corresponding to different rank forms
is zero and the inner product of arbitrary states is obtained
by linearity.
We note here that using this inner product implies that
$P_p^\dagger=P_{\bar p}-i
{\gamma^{\bar q}}_{{\bar p}{\bar q}}$.

The supersymmetry charges are given by
$$
\eqalign{
Q&=\lambda^p\pi_p\cr
Q^*&=Q^\dagger=\lambda^{\bar p}\pi_{\bar p}\cr}
\eqn\susychges
$$
where we have defined
$$
\eqalign{
\pi_p&=P_p+i\lambda^{\bar B}\lambda^A\omega_{pA{\bar B}}\cr
\pi_{\bar p}&=P_{\bar p}+i\lambda^{\bar B}\lambda^A
\omega_{{\bar p}A{\bar B}}\qquad.\cr}
\eqn\a
$$
In defining the quantum supersymmetry charges
we have made a definite choice of operator ordering. We argue in the
appendix that this is consistent with a natural operator ordering
choice in the full field theory.

To see how the supercharges act on
the states, we first present the following useful
commutation  relations:
$$
\eqalign{
\{\lambda^p,\lambda^{\bar q}\}&=\CG^{p\bar q}\cr
[\pi_p,\lambda^q]&=i{\gamma^q}_{pr}\lambda^r\qquad and \qquad c.c.\cr
[\pi_p,\pi_{\bar q}]&=-R_{{\bar r}sp{\bar q}}\lambda^{\bar r}\lambda
^s\qquad.\cr}
\eqn\b
$$
Using these it is straightforward to show that acting on the states \states,
$\pi_p=-i\nabla_p$ and $\pi_{\bar p}=-i\nabla_{\bar p}$.
A short calculation then shows
that the supersymmetry charges act on the states
as follows
$$
\eqalign{
Q^*\vert f>
&=\lambda^{{\bar p}_1}\dots\lambda^{{\bar p}_{p+1}}\vac
{-i\over p!}\nabla_{[{\bar p}_1}
f_{{\bar p}_2\dots{\bar p}_{p+1}]}(X)\qquad.\cr
Q\vert f>
&=\lambda^{{\bar p}_1}\dots\lambda^{{\bar p}_{p-1}}\vac
{-i\over (p-1)!}\nabla^{\bar q}f_{{\bar q}{\bar p}_1
\dots{\bar p}_{p-1}}(X)\qquad.\cr}
\eqn\states
$$
Thus we identify the supersymmetry charges as the anti-holomorphic
exterior derivative and its adjoint:
$$
\eqalign{
Q^*&=-i{\bar\p}\cr
Q&=i{\bar\p}^\dagger\cr}
\eqn\c
$$
where the adjoint is defined with respect to the inner product \inner\
(and thus is strictly only defined on normalisable states).

The quantum Hamiltonian is calculated using the supersymmetry algebra:
$$ \eqalign{ H&=\{Q,Q^*\}\cr &={\bar\p}^\dagger
{\bar\p}+{\bar\p}{\bar\p}^\dagger\cr
&=\CG^{p{\bar q}}\pi_p\pi_{\bar q}+
R_{p{\bar q}}\lambda^{\bar q}\lambda^p\cr}
\eqn\ham
$$
On a K\"ahler manifold
${\bar\p}^\dagger{\bar\p}+{\bar\p}{\bar\p}^\dagger=$$
{1\over 2}(dd^\dagger+d^\dagger d)$
is simply half the Laplacian acting on
differential forms.
The quantum problem can now be tackled in the usual manner. We look
for energy eigenstates of the Hamiltonian \ham, interpreting
the normalisable
wave functions as bound states and the non-normalisable
ones as scattering states.

We first discuss the bound states.
As we have noted a quantum state of the effective action corresponds
to a state
of the full sigma model field theory assuming that all but the zero modes
are in their ground states. Hence, the bound states of the effective
theory with lowest energy
must correspond to bound states in the full theory as there are no states
with lower energy into which they could decay.
The other bound states could correspond to bound states
of the full theory or possibly resonances,
determining which seems a difficult problem.

Thus the normalisable states of lowest energy are an important class of states.
Because the effective action is supersymmetric,
the energy of the states are all greater than zero.
If a state has zero energy then it must be annihilated by all the
supersymmetry charges. In the present case this is equivalent to
the state corresponding to a normalisable harmonic $(0,p)$ form
on the moduli space $\tilde M_N^0$. That is, a
non-trivial element $\bar\p$-cohomolgy
class of the moduli space $\tilde M_N^0$
with the proviso that the state is normalisable.
This proviso is neccessary since the space is non-compact.

We note that the spin of these states could be
ascertained by constructing the angular momentum
operator. The rotation invariance of the parent sigma model implies
that the metric on $\tilde M_N^0$ is $SO(2)$ invariant. Thus there
is a corresponding angular momentum operator of the effective theory
that commutes with the Hamiltonian. Since the bound states have zero
energy and hence zero orbital angular momentum, the angular momentum
of the states is just the spin of the states.

The next stage in the
analysis of the low-energy dynamics of the supersymmetric lumps would be
to investigate the scattering states. This problem seems to require
a rather detailed knowledge of the geometry of the moduli space.
We note that one of the difficulties in pursuing
the quantum mechanics further is that it will depend on which
moduli space $\tilde M_N$ one is considering. Depending
on how one fixes the moduli that are not allowed to vary in time
one will obtain different geometries and possibly topologies
and hence different quantum systems.
It would be wise to
investigate the quantum scattering in the purely bosonic case first,
which has not yet
been attempted.

\chapter{Conclusions}
In this paper we have initiated the investigation of the low-energy
dynamics of solitons of supersymmetric field theories by presenting some
detailed calculations within the context of a class of
N=2 non-linear sigma
models. We developed a number of techniques that can be applied to
many other models. We conclude by summarising the important points.
Generically, the solitons break half of the
supersymmetry and the bosonic and fermionic zero modes form a
multiplet with respect to the unbroken supersymmetry.
This can be used to carry out a supersymmetric collective co-ordinate
expansion and the low-energy effective action will be that of supersymmetric
quantum mechanics based on the moduli space of static bosonic soliton
solutions.
The operator ordering ambiguities in the quantum theory
can be reduced by demanding
that the operator ordering in the soliton sectors be the same as
that in the vaccuum sector.
It seems likely that the Hilbert space of states will be isomorphic to
some kind of differential forms on the moduli space.
Bound states of zero energy will then correspond to normalisable harmonic
forms on the
moduli space and correspond to bound states in the spectrum of the
parent field theory.
Detailed calculations of the scattering theory of the
effective action is an issue that we hope to report on in the future.

\ack
{I would like to thank Peter Bowcock, Peter Freund, Dan Waldram
and in particular
Jeff Harvey for valuable discussions
and criticisms.
This work is supported by a grant from the Mathematical
Discipline Center of the Department of Mathematics, University
of Chicago.}

\appendix
In general there are operator ordering ambiguities
in the quantisation of any classical system.
Demanding that
observables be self adjoint operators and that the Hilbert
space furnish a representation of any symmetry groups present
constrains the operator ordering.
In the semi-classical
quantisation of
theories with solitons a further constraint is that one must
ensure that the operator ordering in the vaccuum sector
is the same as that in the
soliton sectors [\raj].

We will argue at a formal level at least
that the operator ordering chosen
in section 4 is consistent with a natural operator ordering in the
canonical quantisation of the full
supersymmetric sigma model \susysig. The operator ordering of the sigma model
that we will
use agrees with that chosen by Witten in [\wittent] to calculate the Witten
index of the model.

Using the components of the spinor field $\psi^i_\alpha(x)$ we first define
$$
\psi^i={1\over \sqrt 2}(\psi^i_1+i\psi^i_2)
\qquad
\psi^{*i}={1\over \sqrt 2}(\psi^i_1-i\psi^i_2)\qquad
\qquad.
\eqn\bad
$$
We next introduce an orthonormal basis of one forms on the target
manifold of the sigma model $\{e^P_j(\phi(x)\}$ and a spin connection
defined by
$$
{{\omega_i}^P}_Q(\phi(x))=e^P_j(\p_iE^j_Q+{\Gamma^j}_{ik}E^k_Q)
\eqn\fred
$$
where we have also introduced the inverse matrix $E^i_P(\phi(x)$.
In terms of these variables the supersymmetric sigma model action
\susysig\ can be written in the form
$$
S=-{1\over 2}\int d^3x\left\{g_{ij}\partial_m\phi^i\partial
_n\phi^j\eta^{mn}
+i\psi^P\Ds\psi^{*Q}\delta _{PQ}+{1\over 2}R_{PQRS}\psi^{*P}\psi^{^*Q}
\psi^R\psi^S\right\}\eqn\susysigt
$$
where we have defined $\psi^P=e^P_i\psi^i$.

Considering \susysigt\ as a supersymmetric quantum mechanics based
on the infinite dimensional target $\CC$, the canonical quantisation
is now similar to
the quantisation of
the supersymmetric quantum mechanics discussed in section 4. Without presenting
all the details, the non-vanishing equal time commutation
relations are given by
$$
\eqalign{
[\phi^i(x),P_j(y)]&=i\delta^i_j\delta(x-y)\cr
\{\psi^P(x),\psi^{*Q}(y)\}&=\delta^{PQ}\delta(x-y)\cr}
\eqn\etcc
$$
where $P_i(x)$ is the momentum conjugate to $\phi^i(x)$. Defining
a state $\vac$ satisfying
$$
\psi^{i}(x)\vac=0
\eqn\vaccuum
$$
the Hilbert space consists of states of the form
$$
\int dx^{i_1}\dots dx^{i_n}
dF[\phi]_{i_1\dots i_n}(x^{i_1}\dots x^{i_n})
\psi^{i_1}(x^{i_1})\dots\psi^{i_n}(x^{i_n})\vac
\eqn\statey
$$
which we can formally identify with the differential forms on $\CC$.
The momentum operator is realised on these states as a functional
derivative: $P_i(x)=-i{\delta\over \delta\phi^i(x)}$.

The components of the supersymmetry charges are given by
$$
\eqalign{
\tilde Q&=\int d^2x \psi^i\Pi_i\qquad\qquad\qquad
\tilde Q^*=\int d^2x\psi^{*i}\Pi_i\cr
S&=\int d^2x {J_i}^j(\phi(x))\psi^i\Pi_j\qquad
S^*=\int d^2x{J_i}^j(\phi(x))\psi^{*i}\Pi_j\cr}
\eqn\bus
$$
where we have defined
$$
\Pi_i(x)=P_i(x)+i\omega_{iPQ}\psi^{*Q}\psi^P
\eqn\car
$$
In [\wittent] the quantisation of this model was considered in the
zero momentum sector (fields independent of $x$) in order to
calculate the Witten index of the model. It was shown
that with a particular choice of
operator ordering the quantisation led to
a Hilbert space of states isomorphic to the differential forms on the
target manifold of the sigma model and that the supersymmetry charges
acted as the exterior derivative and its adjoint.
It can be checked that the operator ordering we have chosen
is consistent with that chosen in [\wittent].

Using the supersymmetry algebra to calculate the Hamiltonian we
find
$$
2H=\{\tilde Q,\tilde Q^*\}
=\int d^2x\left({1\over \sqrt g}\Pi_i{\sqrt g}g^{ij}
\Pi_j + R_{lkij}\psi^{*i}\psi^j\psi^{*k}\psi^l\right)
\eqn\x
$$
where we have used heavily the K\"ahler propert of $\CC$.
We now want to truncate the theory to the zero-mode sector.
We view the collective co-ordinate expansion \ansatz\ as
change of co-ordinates on $\CC$ from the field $\phi^i(x)$
to its infinite mode expansion. Ignoring all but the collective
co-ordinates will give us the desired operator ordering in
the lump sectors. We first consider the kinetic term in \x.
Formally it is the Laplacian acting on the
differential forms on $\CC$. Since this is independent of the
choice of co-ordinates on $\CC$ it truncates to the Laplacian
acting on the differential forms on $\tilde M_N$.
Thus after substituting \ansatz\ into the curvature term in \x\
we obtain the following truncated Hamiltonian
$$
\eqalign{
H&={1\over 2}[{1\over {\sqrt{\CG}}}\pi^a{\sqrt{\CG}}\CG^{ab}\pi_b
-R_{{\bar p}q{\bar r}s}\lambda^{\bar p}\lambda^q
\lambda^{\bar r}\lambda^s]\cr
&=\CG^{p{\bar q}}\pi_p\pi_{\bar q}+R_{p{\bar q}}\lambda^{\bar q}
\lambda^p\qquad.\cr}
\eqn\hamt
$$
Comparing with \ham\ we verify that the operator orderings agree.

\refout
\end
\bye